# Crossover of Three-Dimensional Topological Insulator of $Bi_2Se_3$ to the Two-Dimensional Limit

Yi Zhang[1]*, Ke He[1]*, Cui-Zu Chang[1,2], Can-Li Song[1,2], Li-Li Wang[1], Xi Chen[2], Jin-Feng Jia[2], Zhong Fang[1], Xi Dai[1], Wen-Yu Shan[3], Shun-Qing Shen[3], Qian Niu[4], Xiao-Liang Qi[5], Shou-Cheng Zhang[5], Xu-Cun Ma[1], and Qi-Kun Xue[1,2]*

[1] Institute of Physics, Chinese Academy of Sciences, Beijing 100190, China

[2] Department of Physics, Tsinghua University, Beijing 100084, China

[3] Department of Physics and Center for Theoretical and Computational Physics, The University of Hong Kong, Pokfulam Road, Hong Kong, China

[4] Department of Physics, The University of Texas, Austin, Texas 78712-0264, USA

[5] Department of Physics, Stanford University, Stanford, California 94305-4045, USA

*e-mail: kehe@aphy.iphy.ac.cn; qkxue@mail.tsinghua.edu.cn

**$Bi_2Se_3$ is theoretically predicted[1,2] and experimentally observed[2,3] to be a three dimensional topological insulator. For possible applications, it is important to understand the electronic structure of the planar device. In this work, thickness dependent band structure of molecular beam epitaxy grown ultrathin films of $Bi_2Se_3$ is investigated by *in situ* angle-resolved photoemission spectroscopy. An energy gap is observed for the first time in the topologically protected metallic surface states of bulk $Bi_2Se_3$ below the thickness of six quintuple layers, due to the coupling between the surface states from two opposite surfaces of the $Bi_2Se_3$ film. The gapped surface states exhibit sizable Rashba-type spin-orbit splitting, due to breaking of structural inversion symmetry induced by SiC substrate. The spin-splitting can be controlled by tuning the potential difference between the two surfaces.**

Dimensionality determines some of the most fundamental properties of a matter, for example, phase transition and correlation[4]. Physics in low-dimensional materials and in crossover from three-dimensional (3D) to two-dimensional (2D) materials is among the most studied topics in condensed matter physics, especially due to the rapid development of advanced growth and characterization techniques with accuracy of atomic layers[5]. Very recently, a new class of matters, topological insulators (TI), was theoretically discovered[1,2,3,6-8] and experimentally demonstrated[2,3,9-13]. TI has gapless and spin-split surface/edge states which are protected by bulk topological properties and therefore robust against disorder and contamination[6-8]. Because of these unique characters, this new kind of matter is not only of great significance in fundamental condensed matter physics, but also promising for applications in spintronics and topological quantum computation[14-16]. TIs at different dimensionalities exhibit more exotic effects. 2D TI, the quantum spin Hall (QSH) phase[6], is characterized by the 1D gapless edge states where opposite spin states counter-propagate. On the other hand, 3D TI is characterized by a surface state consisting of a single Dirac cone, where the spin points perpendicular to the momentum. The finite-size effects of both 2D[17] and 3D[18-20] TIs were studied theoretically within the effective models of Refs 1 and 6, and by first-principle calculations. Due to the coupling between the surface states, a finite energy gap is predicted to open in the 2D limit of the 3D topological insulators. Furthermore, the 2D energy gap is predicted to oscillate between the ordinary insulator gap and QSH gap as a function of thickness. Recently, thin films of $Bi_2Se_3$ have been grown by molecular beam epitaxy (MBE)[21], and by vapor-liquid-solid in nano-ribbon form[22]. For possible applications of the topological insulators, planar devices would be most promising. In this work, by combining MBE, angle-resolved photoemission spectroscopy (ARPES) and model calculations, we have systematically studied the band structure of 3D TI $Bi_2Se_3$ grown by MBE with a thickness from one quintuple layer (QL) to 50 QL. Our results reveal how the electronic and spin structures of a TI evolve during dimensional crossover.

Figures 1a and 1b show the reflection high electron energy diffraction (RHEED) pattern and STM image of a typical MBE-grown $Bi_2Se_3$ film with a thickness of 50 QL, respectively. The sharp 1×1 streaks and atomically flat morphology demonstrate the high crystal quality of the film. Figure 1c shows the band map of the film. The bands show similar structure and dispersion to bulk $Bi_2Se_3$[2]. Two bands with nearly linear dispersion cross with each other at $\bar{\Gamma}$ point, forming a Dirac cone. They are the gapless surface states of $Bi_2Se_3$. The band map shown in Fig. 1c does not exhibit the conduction band like in the bulk samples[2]. The Dirac point is located at 0.12 eV below the Fermi level. The observation indicates that our MBE-grown $Bi_2Se_3$ samples are an nearly intrinsic TI with a single Dirac point and contain fewer impurities, compared to the bulk samples[2,23].

By delicate control of the growth condition, we were able to achieve QL-by-QL growth of the $Bi_2Se_3$ thin films. Figure 1d shows a series of normal emission spectra of $Bi_2Se_3$ films from 1 QL to 6 QL with a step of ~0.3 QL. The peaks indicated by bars are QWSs of the valence band, which will be addressed later. With increasing thickness, stepwise change in the QWS peaks is clearly observed in Fig. 1d, which indicates QL-by-QL growth mode[24]. It enables us investigate the thickness-dependent band structures of the $Bi_2Se_3$ films with an accuracy of QL.

Figures 2a-2e display a series of ARPES band maps of the $Bi_2Se_3$ films from 1 QL to 6 QL (except for 4 QL) measured at room temperature (RT). All the spectra are taken along $\bar{\Gamma}-\bar{K}$ direction. The spectra of 1 QL (Fig. 2a) exhibit a nearly parabolic band dispersion towards the Fermi level along $k_{//}$. At 2 QL (Fig. 2b), the band moves down by 40 meV and another band appears below it. As shown later, the two bands finally evolve into the gapless surface states in thicker films. We refer them hereafter as the upper and lower surface states, respectively. At 3 QL (Fig. 2c), the gap between the upper and lower surface states decreases. Interestingly, the upper surface states are split into two branches. The splitting is more prominent at larger wave vectors but disappears at $\bar{\Gamma}$ point as shown in the energy distribution curves (EDCs) in Fig. 2f. This is a Rashba-type splitting, in which the two sub-bands with different spins shift

along $k_{//}$ axis in opposite directions and degenerate at $\bar{\Gamma}$ point. For thicker films at 5 QL and 6 QL (Figs. 2d and 2e), the gap between the two surface states decreases further. Meanwhile, the outer two branches of the split upper surface states fade out gradually. At 6 QL, the gap disappears and so does the outer branches of the upper surface states. The inner branches of the upper surface states connect the lower surface states, resulting in a Dirac cone (Fig. 1c).

The Dirac point at 6 QL is located at -0.26 eV, lower by 0.14 eV than that of thick (50 QL) $Bi_2Se_3$ films shown in Fig. 1c. It suggests that the film is electron-doped. Above 6 QL, Dirac point moves towards the Fermi level with increasing thickness until ~20 QL where the saturation value is reached. Figure 3a plots the thickness dependence of the Dirac point position. For comparison, the measured mid-gap energies of the surface states and the spin-degenerate points of the Rashba-split upper surface states are also shown. Since STM experiments show little difference in surface defect density between thick and thin films, the thickness dependent electron-doping may come from band bending due to charge transfer from the SiC substrate.

Besides the surface states, above 2 QL, there are other features either above the upper surface states or below the lower surface states. With the increasing thickness, these bands move up or down towards the surface states, and more bands emerge. In Fig. 3b we summarize the energies of these bands at $\bar{\Gamma}$ point with respect to the mid-gap energies of the surface states. The mid-gap energies are used as reference to remove the influence of chemical potential change. The energy evolution shows a typical QWS behavior. Therefore, these bands can be attributed to QWSs of conduction and valence bands.

When the thickness of a film is reduced to only several nanometers, overlapping between the surface state wave functions from the two surfaces of the film becomes non-negligible, and hybridization between them has to be taken into account. For a freestanding and symmetric $Bi_2Se_3$ film, the spin-polarized surface states at one surface will be mixed up with the components of opposite spin from the other surface when the thickness is small enough. It will lead to a hybridization gap at Dirac point to avoid crossing of bands with the same quantum numbers. The observed gap opening below 6 QL can preliminarily be attributed to this effect. The band and spin structures of a 3D TI like $Bi_2Se_3$ near the Fermi level are well described by the four-band effective model[1]. In the ultra-thin limit[18], the 3D TI model of Ref. 1 reduces to the Bernevig-Hughes-Zhang (BHZ) model[6] of the 2D QSH insulator, and the in-plane dispersions of the surface states can be written as[18]:

$$E_{\pm}(k_{//}) = E_0 - Dk_{//}^2 \pm \sqrt{(v_F \hbar k_{//})^2 + (\frac{\Delta}{2} - Bk_{//}^2)^2} \quad . \tag{1}$$

The term including Fermi velocity $v_F$ is the Dirac component of the bands, and $\Delta$ represents the gap size. For 1 QL film, only one band can be observed. The absence of the lower surface states may be due to the bonding with the substrate. The ARPES spectra of 2 QL can be well fitted by Eq. (1). The pink dashed lines in Fig. 2b depict the fitted curves with the parameters listed in Table I. However, starting from 3 QL, the surface states exhibit Rashba-type splitting. The splitting cannot be obtained with Eq. (1), and neither by previous first-principle calculation[1]. Both the effective model and the first-principle calculation are based on a free-standing symmetric TI film, whereas the existence of substrate breaks the structural inversion symmetry, which can result in a Rashba-type spin splitting in the ultrathin film regime.

To model the substrate effect, we invoke an effective electrical potential $V(z)$ along the film normal direction. From this modified model, we can obtain the dispersions[25]:

$$E_{\sigma\pm}(k_{//}) = E_0 - Dk_{//}^2 \pm \sqrt{(|\tilde{V}'| + \sigma v_F \hbar k_{//})^2 + (\frac{\Delta}{2} - Bk_{//}^2)^2} \quad . \tag{2}$$

The $\tilde{V}'$ term represents the effect of the potential $V(z)$. Each spin-degenerate dispersion in Eq. (1) is now split into two sub-bands of opposite spins ($\sigma = \pm 1$) which shift away from each other along $k_{//}$ axis. If $\tilde{V}' = 0$, Eq. (2) degenerates into Eq. (1), which corresponds to the 2 QL case discussed above. The splitting of the upper surface states is very well reproduced in the fitting with Eq. (2) (see the dashed lines in Figs. 2c-2e). When $\Delta$ is zero, the dispersions obtained from Eq. (2) evolve into two sets of Dirac cones centered at $\bar{\Gamma}$. The two Dirac points are shifted in energy by $2|\widetilde{V}'|$. In ARPES data, at 6 QL only one Dirac cone was observed. The outer branches of the Rashba-split upper surface states, which are expected to evolve into part of the other Dirac cone, are nearly invisible. In Fig. 3c, the calculated weights in real space of the surface states at 5 QL are shown with the colours of the lines. It is clear that near critical thickness the outer and inner branches of the Rashba split surface states localize at different surfaces of the film. The two Dirac cones obtained from Eq. (2) correspond to the two sets of gapless surface states from the surface and interface sides of the film, respectively. The absence of one Dirac cone in ARPES spectra is due to its localization at the interface side, which is beyond the detection depth of ARPES. Under this context, the observed Rashba-splitting can also be understood as a result of the hybridization of the two sets of gapless surface states with different Dirac points from the two surfaces of the film (see the inset of Fig. 3c).

The breaking of structural inversion symmetry of the system can originate from the band bending induced by the substrate, as suggested in Fig. 3a, or alternatively from different environments of the two surfaces. In the former case, when the film is much thinner than the thickness of band-bending region, i.e. Debye length, which is determined by the dielectric constant and carrier density of the material, the potential variation along the z direction of the film will be negligible. So Rashba effect is weaker in thinner films. In the latter case, structural asymmetry is more significant in thinner films, and thus Rashba effect is stronger. Our observation of reduced Rashba splitting with decreasing film thickness suggests that the former case dominates in $Bi_2Se_2$ films. In an epitaxial film, band bending can be modified by surface photovoltage (SPV) effect, which takes place in a p-n or Schottky junction in which the photon excited electrons and holes are driven to opposite directions by a space charge layer[26,27]. As a result, the band bending is reduced and even completely removed. SPV effect is more prominent at low temperature (LT) due to less electron-hole recombination. The epitaxial $Bi_2Se_3$ films show a strong SPV effect at 150K, which is confirmed by the 0.06 eV shift of the Fermi edge compared with the RT spectra. In Fig. 3e, we show EDCs of ARPES spectra of 3 QL $Bi_2Se_3$ at 150 K. Rashba splitting cannot be distinguished any more, while other features, for example the gap size, are basically similar to the RT data. The thickness dependence of energies of the Dirac point (≥6 QL) and mid-gap position (<6 QL) measured at LT is also shown in Fig. 3a. The variation of chemical potential with thickness is largely reduced, suggesting a nearly flat band. The quench of Rashba splitting by SPV effect demonstrates that the splitting is mainly contributed by substrate-induced band bending.

The above conclusion implies that the Rashba splitting can be controlled by modifying band bending, which can be easily realized by applying a gate voltage. It is of great significance for spintronics devices, especially Datta-Das spin field-effect transistor[28]. In Table I, the fitted Rashba parameter $\alpha_R$ values are listed. The $Bi_2Se_3$ films (3 QL-6 QL) show at least one order of magnitude larger Rashba splitting than semiconductor heterostructures[29], and can be removed by changing the band bending by ~0.3 eV. Therefore ultrathin films of topological insulators provide a new kind of spintronics materials that may realize high efficiency electrical spin manipulation.

Thickness-dependent oscillatory transition with a periodicity of 3QL between QSH and ordinary insulator phases was predicted theoretically[19]. The oscillation mostly refers to the sign of the gap, but could also lead to an oscillation in the magnitude of the gap size[18-20]. In Fig. 3d, we show the measured gap size as a function of thickness, which is more or less monotonic. With the ARPES result itself, which only measures the magnitude but not the sign of the gap, it is difficult to judge if the films are in ordinary insulator or QSH phases. Further experiments by STM and electrical transport measurement with and without external magnetic field, similar to those carried out in Ref. 9, are needed to clarify this issue.

In summary, thickness-dependent ARPES study reveals gap-opening and Rashba spin-orbit splitting in the topologically protected surface states of ultrathin $Bi_2Se_3$ films. It results from the coupling between the surface states from the opposite surfaces with different chemical potentials. The existence of gapless surface states at interface side of a TI provides an evidence for the robustness of topological states.

**Methods**

All experiments were performed in an ultra-high vacuum system (Omicron), equipped with MBE, scanning tunneling microscope (STM) and ARPES. The base pressure of the system is $1.5 \times 10^{-10}$ Torr. In ARPES measurement, photoelectrons are excited by an unpolarized He-Iα light (21.21 eV), and collected by a Scienta SES-2002 analyzer (15 meV). $Bi_2Se_3$ films were grown under Se-rich conditions on a double-layer graphene terminated 6H-SiC (0001) substrate at 220 °C by MBE. Bi (99.9999%) and Se (99.999%) were both evaporated from standard Knudsen cells. The growth rate was calibrated by real time RHEED intensity oscillation measured on the (00) diffraction. The $Se_4(Se_2)$/Bi flux ratio was between 10 and 15, which leads to a growth rate of ~0.3 QL/min when the Bi source temperature was set at 550 °C.

**Acknowledgements**

This work was supported by National Science Foundation and Ministry of Science and Technology of China and RGC of Hong Kong, China under grant No. HKU 7037/08P. X.L.Q. and S.C.Z. are supported by the Department of Energy, Office of Basic Energy Sciences, Division of Materials Sciences and Engineering, under contract DE-AC02-76SF00515.

**Author contributions**

K.H., S.Q.S., X.C.M., and Q.K.X. conceived and designed the experiments. Y.Z., K.H., and C.Z.C. carried out MBE growth and ARPES measurements. C.L.S. carried out substrate preparation and STM observation. L.L.W., X.C., J.F.J., X.C.M., and Q.K.X. assisted the experiments. W.Y.S. and S.Q.S. carried out the theoretical analyses and data interpretations. Q.N., Z.F., X.D., X.Q., and S.C.Z. assisted the theoretic analyses. Y.Z., K.H., S.Q.S., S.C.Z., and Q.K.X. prepared the manuscript.

**Additional information**

Correspondence and requests for materials should be addressed to Q.K.X. o K.H.

**Figure Legends**

**Figure 1 | Growth of $Bi_2Se_3$ films.** **a**, RHEED pattern of a 50 QL $Bi_2Se_3$ film grown by MBE. **b**, STM image of the film. **c**, ARPES spectra of the film along $\overline{\Gamma}-\overline{K}$ direction. **d**, Thickness-dependent normal emission photoemission spectra of the $Bi_2Se_3$ films at different thicknesses. The thickness step is ~0.3 QL.

**Figure 2 | ARPES spectra of $Bi_2Se_3$ films at room temperature.** **a-e**, ARPES spectra of 1 QL, 2 QL, 3 QL, 5 QL, and 6 QL along $\overline{\Gamma}-\overline{K}$ direction measured at RT. **f**, Energy distribution curves (EDCs) of **c**. **g**, EDCs of **d**. **h**, EDCs of **e**. The pink dashed lines in **b** represent the fitted curves using Eq. (1). The blue and red dashed lines in **c**-**e** represent the fitted curves using Eq. (2). The corresponding fitting parameters are listed in Table I.

**Figure 3 | Analyses of the ARPES data of Fig. 2** **a**, Thickness dependence of the binding energies of the Dirac point (≥ 6 QL) (hollow black squares), the spin-degenerate point (<6 QL) (hollow blue triangles), and the mid-gap position $E_0$ of the surface states (<6 QL) measured at room temperature (hollow green circles) and 150K (solid red circles). **b**, Thickness dependent energies of QWS peaks at $\overline{\Gamma}$ point relative to mid-gap energy $E_0$. **c**, Real-space weight of the surface states at 5 QL. Green and blue colours represent the states that mainly localize at surface and interface, respectively. The inset shows a schematic illustration of the surface states of $Bi_2Se_3$ film above (right) and below (left) 6 QL. The solid and dashed lines represent the surface states that mainly localize at surface and interface of $Bi_2Se_3$ film, respectively. The red and blue colours of the lines represent different spins. **d**, Thickness dependence of the measured gap size of the surface states. **e**, ARPES EDCs of the 3 QL $Bi_2Se_3$ film measured at 150 K.

**TABLE I.** Parameters of Eqs (1) and (2) used to fit the bands in Figs. 2b-2e, and the fitted Rashba parameters ($\alpha_R$).

| QL | $E_0$ (eV) | $D$ (eV·Å²) | $\Delta$ (eV) | $B$ (eV·Å²) | $v_F$ (10⁵ m/s) | $|\tilde{V}'|$ (eV) | $\alpha_R$ (eV·Å) |
|---|---|---|---|---|---|---|---|
| 2 | -0.470 | -14.4 | 0.252 | 21.8 | 4.71 | 0 | 0 |
| 3 | -0.407 | -9.7 | 0.138 | 18.0 | 4.81 | 0.038 | 0.71 |
| 4 | -0.363 | -8.0 | 0.070 | 10.0 | 4.48 | 0.053 | 1.27 |
| 5 | -0.345 | -15.3 | 0.041 | 5.0 | 4.53 | 0.057 | 2.42 |
| 6 | -0.324 | -13.0 | 0 | 0 | 4.52 | 0.068 | 2.78 |

Figure 1

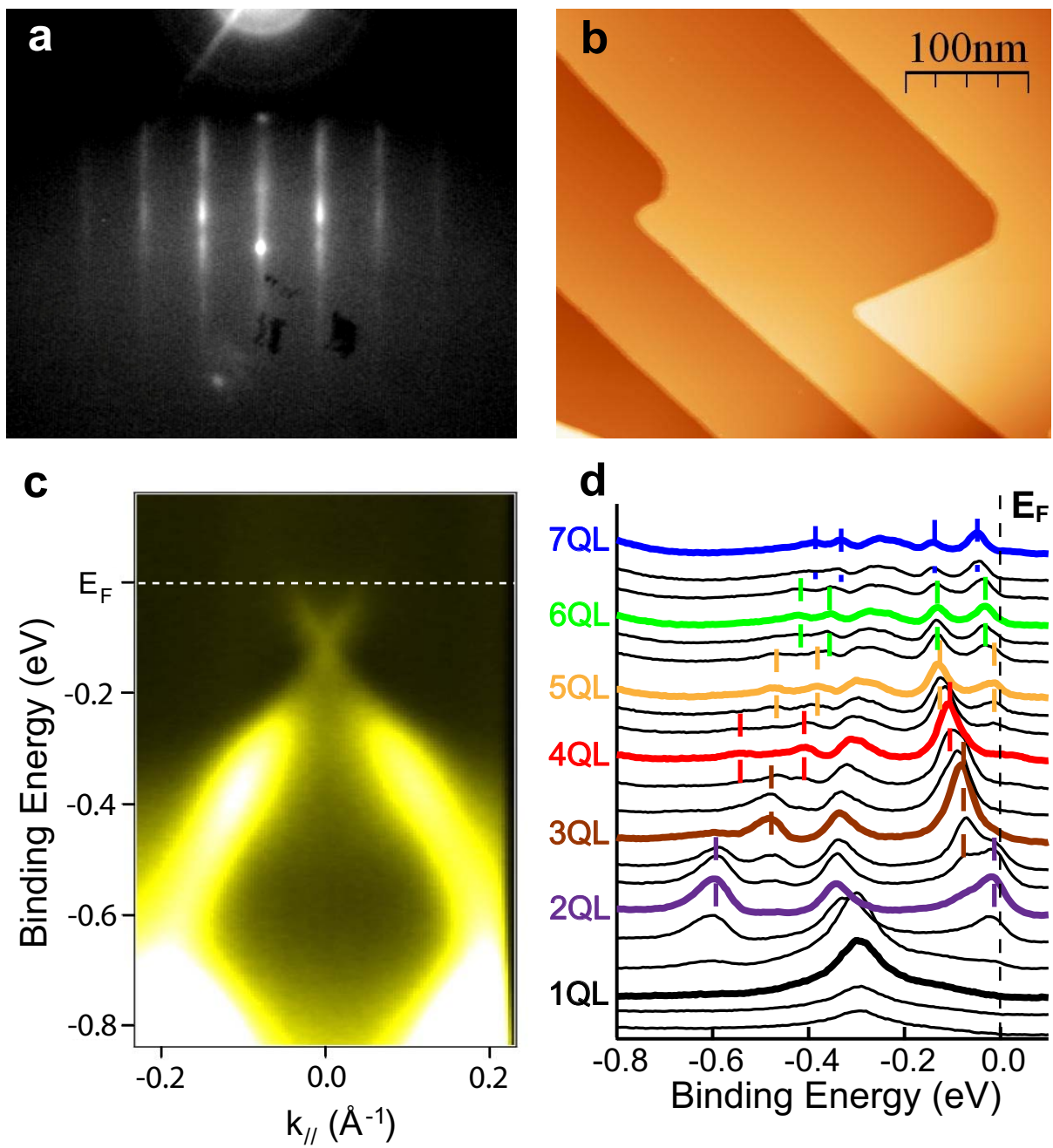

Figure 2

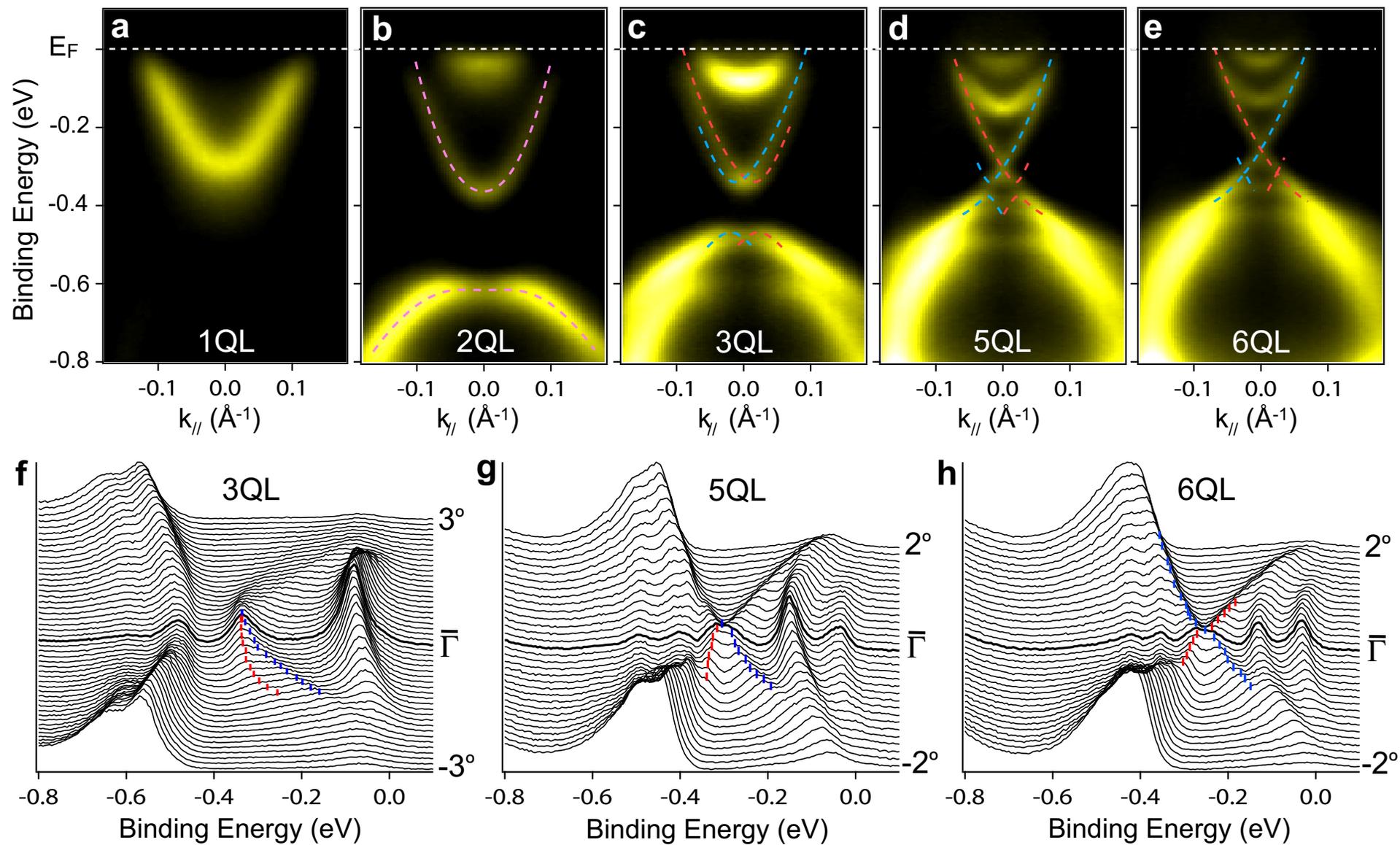

Figure 3

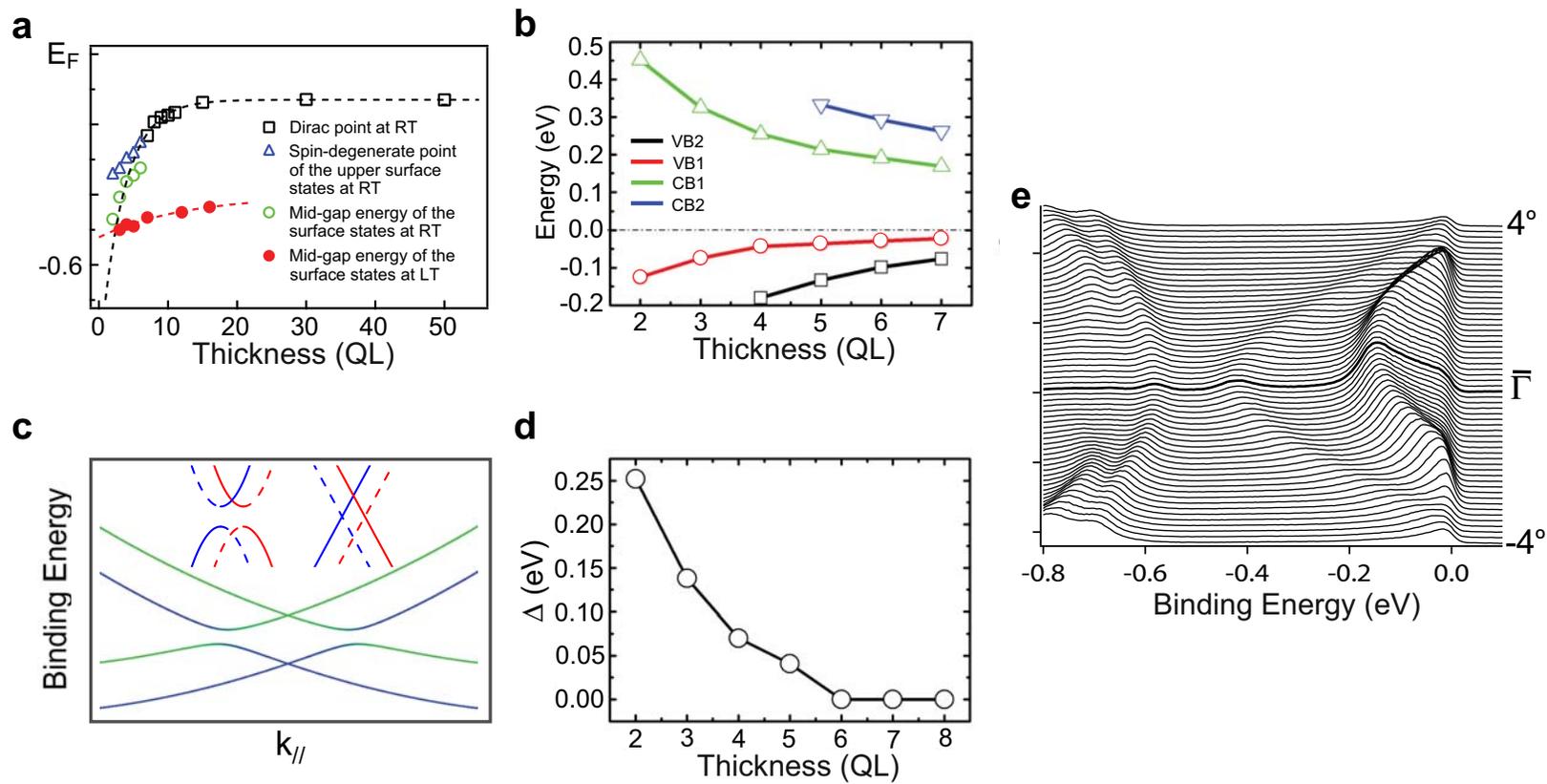